\documentclass[runningheads]{llncs}

\usepackage[T1]{fontenc}
\usepackage{graphicx}
\usepackage[hyphens]{url}
\usepackage[colorlinks=true, allcolors=black]{hyperref}
\usepackage{subfig}
\usepackage{subcaption}
\usepackage{enumitem}
\usepackage{xcolor}
\usepackage{biblatex}
\usepackage{array}  
\usepackage{amsmath}
\usepackage{float}
\restylefloat{table}
\usepackage{booktabs}
\usepackage[justification=centering]{caption}
\usepackage{placeins}
\usepackage{algorithm,algcompatible,amsmath}

\begin{document}
\title{Advanced Encryption Technique for Multimedia Data Using Sudoku-Based Algorithms for Enhanced Security}

% \author{
% Mithil Bavishi\orcidID{0009-0004-3526-8813} \and
% Anuj Bohra\orcidID{0009-0005-0532-0745} \and
% Kushal Vadodaria\orcidID{0009-0007-3278-2077} \and
% Abhinav Bohra\orcidID{0009-0008-8377-0604} \and
% Neha Katre\orcidID{0000-0001-8320-7071} \and
% Ramchandra Mangrulkar\orcidID{0000-0002-9020-0713} \and
% Vinaya Sawant\orcidID{0000-0003-4661-9231}
% }

% \titlerunning{Advanced Encryption Technique for Multimedia} 
% %
% \authorrunning{Mithil Bavishi et al.}

% \institute{Dept. of Information Technology, SVKM's Dwarkadas J. Sanghvi College of Engineering, Mumbai-400056, India\\
% \email{bavishimithil@gmail.com, anujbohra232003@gmail.com, kushalvadodaria@gmail.com, neha.katre@djsce.ac.in, ramchandra.mangrulkar@djsce.ac.in, vinaya.sawant@djsce.ac.in}
% Dept. of Information Technology, SVKM's 
% Dwarkadas J. Sanghvi College of Engineering, Mumbai-400056, India\\
% \and
% Assistant Professor, Dept. of Information Technology, SVKM's Dwarkadas J. Sanghvi College of Engineering, Mumbai-400056, India\\
% \email{neha.katre@djsce.ac.in}
% \and 
% Professor, Dept. of Information Technology, SVKM's Dwarkadas J. Sanghvi College of Engineering, Mumbai-400056, India\\
% \email{ramchandra.mangrulkar@djsce.ac.in}

% \maketitle              % typeset the title of the contribution
%

\author{
Mithil Bavishi \and
Anuj Bohra \and
Kushal Vadodaria \and
Abhinav Bohra \and
Neha Katre \and
Ramchandra Mangrulkar \and
Vinaya Sawant
}

\titlerunning{Advanced Encryption Technique for Multimedia}
\authorrunning{Mithil Bavishi et al.}

\institute{
\email{
bavishimithil@gmail.com \\
anujbohra232003@gmail.com \\
kushalvadodaria@gmail.com \\
bohraabhinav8@gmail.com \\
neha.katre@djsce.ac.in \\
ramchandra.mangrulkar@djsce.ac.in \\
vinaya.sawant@djsce.ac.in
}
}

\maketitle

\begin{abstract}
Encryption and Decryption is the process of sending a message in a ciphered way that appears meaningless and could be deciphered using a key for security purposes to avoid data breaches. This paper expands on the previous work on Sudoku-based encryption methods, applying it to other forms of media including images, audio and video. It also enhances the security of key generation and usage by making it dependent on the timestamp of when the message was transmitted. It is a versatile system that works on multimodal data and functions as a block-based transposition cipher. Instead of shuffling, it can also employ substitution methods like XOR, making it a substitution cipher. The resulting media are highly encrypted and resilient to brute-force and differential attacks. For images, NPCR values approach 100\% and for audio, SNR values exceed 60dB. This makes the encrypted audio significantly different from the source, making decryption more difficult.

\keywords{Sudoku-based encryption \and Computational cryptography \and Data security \and Block cipher\and Multi-modal data security}
\end{abstract}
\section{Introduction}

The transfer of multimedia files, such as audio, video, and images, is essential across various sectors, including government, military, and finance. Ensuring the security and privacy of these files during transfer is crucial due to the high risk of data breaches. Encryption techniques play a vital role in protecting this data. It is important to reduce data in a way that results in a tolerable loss of quality. Speech and Video frames can be compressed without significantly sacrificing speech intelligibility by ignoring and discarding tiny and lesser coefficients and data, then utilizing encryption techniques.  

Several key encryption algorithms have been used to transfer media like SHA, RSA and ECC. This paper showcases a new method for multimedia encryption using Sudoku as the encryption and decryption key leverages an NP-complete problem,
ensuring heightened security, which is applied to different media formats including images, audio, and video. By utilizing the complexity of Sudoku puzzles and dynamic key generation based on timestamps, this method enhances security. The proposed algorithm is analyzed and compared with other encryption techniques, demonstrating superior resilience against brute-force and differential attacks. This approach ensures robust encryption for various media types, significantly reducing the risk of unauthorized access and tampering.

\section{Literature review}

In 2023, (Aadhitya et al., ~\hyperref[ref1]{[1]}) proposed HexE, which uses NxN Sudoku puzzles and UNIX timestamps to create a novel method for securing audio content in voice chat. The process renders the encrypted content unintelligible without the proper decryption key, seeking to secure voice communications. The encryption key is generated from a timestamp, and any alteration during the decryption process leads to failure. However, the paper's use of audio shuffling as an encryption method could be deciphered by attackers, which is addressed by integrating padding, shuffling, and rotation to further obfuscate the content.

In 2023, (Deshpande et al., ~\hyperref[ref2]{[2]}) introduced a novel Sudoku-based encryption algorithm for enhancing image security. The method utilizes 9x9 and 16x16 Sudoku matrices for key generation with multiple iterations and modified thresholding to strengthen security. Sudoku is employed to calculate the threshold value for the encryption key. The paper draws inspiration from the work of (Dhruv et al., ~\hyperref[ref3]{[3]}), which introduced a unique key generation technique using a die-roll method for RGB image cryptography. Deshpande et al.'s method is compared with other encryption algorithms for security analysis and has implications for broader multimedia security analysis.

Previously , (Loukhaoukha et al., ~\hyperref[ref4]{[4]}) proposed an efficient image encryption algorithm based on block permutation and the Rubik's cube principle, specifically for iris images. The algorithm employs logical gates, such as XOR, to enhance the encryption process. This cryptosystem offers robust security against major threats. The research results show that incorporating the XOR operator adds extra security to the system, making it resilient against differential and statistical attacks.

In 2021, (Abduljaleel et al., ~\hyperref[ref5]{[5]}) introduced a method for speech signal compression and encryption utilizing Sudoku, fuzzy C-means clustering, and the Threefish cipher for secure transmission and efficient storage. The process involves scrambling the speech signal after frequency analysis, removing less significant components, and then applying the Threefish algorithm for encryption to ensure confidentiality. This approach mirrors the earlier work of (K. Venkatesh et al., ~\hyperref[ref6]{[6]}), where similar techniques for compressing and encrypting speech signals were employed to reduce storage needs and enhance security.

In 2021, (Abed et al., ~\hyperref[ref7]{[7]}) proposed a method to enhance data security and integrity using MinHash combined with RSA and AES algorithms. The study introduces a technique that divides data into shingles, aiding in the development of encryption and decryption keys. The MinHash technique improves data integrity checks by comparing enciphered digests between the sender and receiver. The findings indicate that this approach reduces encryption time and enhances data integrity performance, particularly for large datasets.

A study (Ashwita et al., ~\hyperref[ref8]{[8]}) analyzing various video encryption algorithms, focusing on the use of bitstreams and frames for encryption. The research compares major video compression algorithms, their performance rates, and their resilience against brute-force attacks. A key concern highlighted is the reduction in video quality during transmission. Similarly, (Obaida et al., ~\hyperref[ref9]{[9]}) compared symmetric, chaotic, and asymmetric algorithms, emphasizing the advantages and disadvantages of each. Their review highlights that while full encryption offers robust security, its slow processing speed makes it unsuitable for real-time applications, whereas selective encryption is faster but less secure. The study suggests hybrid algorithms as a potential solution, combining speed and security for effective real-time video encryption.

In 2020, (Saidi et al.,  ~\hyperref[ref10]{[10]}) explored the use of the Number of Pixel Change Rate (NPCR) and Unified Average Changing Intensity (UACI) metrics for sensitivity analysis of encrypted inSAR interferograms generated from satellite data. The study highlights that these metrics are crucial for evaluating the security and robustness of encryption algorithms, particularly in sensitive remote sensing applications. The ability to assess how encryption affects pixel structure and intensity is essential for maintaining data integrity. The findings suggest that NPCR and UACI can help identify the most reliable encryption schemes, contributing to the development of secure encryption methods for earth observation and remote sensing systems.

In 2023, (Veera et al., ~\hyperref[ref11]{[11]}) proposed a technique to enhance information security, focusing on text and image data encryption. The method combines the Card Deck Shuffle Algorithm with a Modified Caesar Cipher to create a two-stage encryption process, making it highly efficient. The use of an \( n + 8 \) bit key results in \( 2^{26} \) possible key combinations, providing a robust encryption system. The algorithm was rigorously tested for key sensitivity and correlation analysis, demonstrating its effectiveness in making it difficult for attackers to guess the encryption key.

\section{Proposed Methodology}
This section describes the algorithm and the encryption pattern used for it. The proposed system employs a multi-staged approach to securely process various media formats, ensuring robust security and dynamic key management while maintaining efficiency across multiple media types.

\subsection{System Design}
The proposed system design utilizes a multi-staged approach to securely process various media formats such as images, audio, and video. It ensures robust security and dynamic key management while maintaining efficiency across multiple media types.

\begin{figure}
    \centering
    \includegraphics[width=1\linewidth]{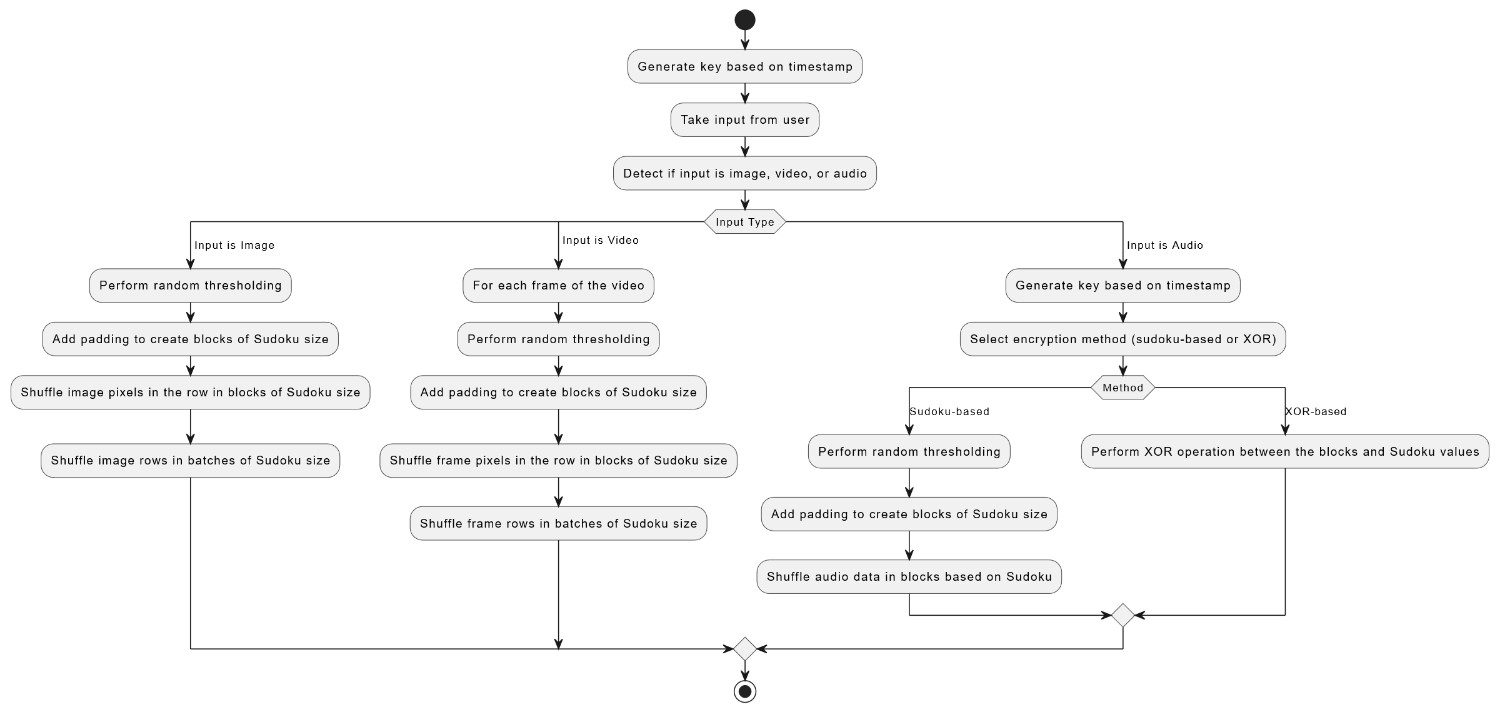}
    \caption{System Flowchart}
    \label{fig:sysarch}
\end{figure}

As shown in Fig. \ref{fig:sysarch}, the flowchart shows that the Sudoku-based encryption algorithm is designed to handle images, audio, and video files securely. It starts with selecting and validating the input files. Then, the encryption process involves several steps: converting the data to a binary format, then adding padding so that Sudoku transformations can be applied to it, shuffling it based on a Sudoku puzzle, applying the Sudoku transformation, and rotating the data. The encryption key is generated dynamically using the current timestamp, ensuring each session has a unique key. Decryption reverses these steps to restore the original data. Finally, the system outputs the encrypted or decrypted files, ready for secure storage or transmission. This approach provides strong security and efficient processing for different types of media.

The use of Sudoku in encryption has the following benefits compared to other encryption techniques. First of all, the use of timestamps in key generation to make every encryption session to have its unique key is very secure as it reduces the chances of reusing the keys. Secondly, the application of Sudoku transformations adds a layer of complexity whereby it is almost impossible to decode the information without the help of the key, which also guards the information against hacker attack. Furthermore, since the algorithm is capable of processing images, audio, as well as video data, it can be considered as a reliable system for protection of multimedia contents. This approach is very helpful for the applications where the data need to be secured and this includes security within communication, distribution of media within computer networks and sensitive information storage so that the sensitive information do not get compromised at any stage. 

\subsection{Algorithms}
In this section, algorithms are outlined that are used for processing and encrypting video, audio and image data. The algorithms include techniques for shuffling and unshuffling pixels in images, encrypting and decrypting image blocks, and applying thresholding to audio data and breaking videos into their frames, encrypting them, finally joining them to create the encrypted video. Each algorithm is designed to handle specific data manipulations.

These algorithms use the inherent complexity of Sudoku puzzles to encrypt data with multiple layers of security. The algorithms used in images make sure that every single pixel block is shuffled and changed, resulting in an output that is nearly hard to predict or undo without the right key. Thresholding techniques are used to alter the amplitude values of audio signals in order to ensure that the encrypted audio produced differs significantly from the original. Frame-by-frame encryption is applied to video data, treating each frame as a separate image that is shuffled, altered, and then put back together to provide complete security for the video file.

\subsubsection{Algorithms for Images}
This section presents an overview of the key algorithms utilized in image processing.

\begin{algorithm}
    \caption{Threshold Image Encryption}
    \label{alg1}
    \begin{algorithmic}[1]
        \REQUIRE image\_path, randomNumber
        \ENSURE encrypted image saved as "threshold\_image.png"
        \STATE img $\leftarrow$ \texttt{Image.open(image\_path)}
        \STATE img\_array $\leftarrow$ \texttt{np.array(img)}
        \IF{\texttt{len(img\_array.shape) == 3}} \COMMENT{Color image}
            \FOR{each $i$ in \texttt{range(img\_array.shape[0])}}
                \FOR{each $j$ in \texttt{range(img\_array.shape[1])}}
                    \FOR{each $k$ in \texttt{range(img\_array.shape[2])}}
                        \IF{img\_array[$i, j, k$] + randomNumber $\leq$ 255}
                            \STATE img\_array[$i, j, k$] $\leftarrow$ img\_array[$i, j, k$] + randomNumber
                        \ELSE
                            \STATE img\_array[$i, j, k$] $\leftarrow$ img\_array[$i, j, k$] - 255 + randomNumber
                        \ENDIF
                    \ENDFOR
                \ENDFOR
            \ENDFOR
        \ELSIF{\texttt{len(img\_array.shape) == 2}} \COMMENT{Grayscale image}
            \FOR{each $i$ in \texttt{range(img\_array.shape[0])}}
                \FOR{each $j$ in \texttt{range(img\_array.shape[1])}}
                    \IF{img\_array[$i, j$] + randomNumber $\leq$ 255}
                        \STATE img\_array[$i, j$] $\leftarrow$ img\_array[$i, j$] + randomNumber
                    \ELSE
                        \STATE img\_array[$i, j$] $\leftarrow$ img\_array[$i, j$] - 255 + randomNumber
                    \ENDIF
                \ENDFOR
            \ENDFOR
        \ENDIF
        \STATE processed\_img $\leftarrow$ \texttt{Image.fromarray(img\_array)}
        \STATE \texttt{processed\_img.save("threshold\_image.png")}
    \end{algorithmic}
\end{algorithm}

\begin{algorithm}
    \caption{Pad and Shuffle Image}
    \label{alg2}
    \begin{algorithmic}[1]
        \REQUIRE image\_path, SudokuSize
        \ENSURE padded and shuffled image saved as "padded\_and\_shuffled\_image.png", seed used for shuffling
        \STATE img $\leftarrow$ \texttt{Image.open(image\_path)}
        \STATE width, height $\leftarrow$ \texttt{img.size}
        \STATE pad\_height $\leftarrow$ \texttt{SudokuSize - height if height < SudokuSize else (height // SudokuSize + 1) * SudokuSize - height if height \% SudokuSize != 0 else 0}
        \STATE pad\_width $\leftarrow$ \texttt{SudokuSize - width if width < SudokuSize else (width // SudokuSize + 1) * SudokuSize - width if width \% SudokuSize != 0 else 0}
        \STATE img $\leftarrow$ \texttt{ImageOps.expand(img, (0, 0, pad\_width, pad\_height))}
        \STATE img\_array $\leftarrow$ \texttt{np.array(img)}
        \STATE seed $\leftarrow$ \texttt{pad\_width + height}
        \IF{\texttt{seed \% 9 != 0}}
            \STATE seed $\leftarrow$ \texttt{seed + 9 - (seed \% 9)}
        \ENDIF
        \STATE \texttt{np.random.seed(seed)}
        \STATE \texttt{np.random.shuffle(img\_array)}
        \STATE img $\leftarrow$ \texttt{Image.fromarray(img\_array)}
        \STATE \texttt{img.save("padded\_and\_shuffled\_image.png")}
    \end{algorithmic}
\end{algorithm}

\begin{algorithm}
    \caption{Sudoku Based Shuffling of Image}
    \label{alg3}
    \begin{algorithmic}[1]
        \REQUIRE image\_path
        \ENSURE shuffled image saved as "shuffled\_image.png"
        \STATE perm $\leftarrow$ \texttt{input from user or keys.txt}
        \STATE perm $\leftarrow$ \texttt{np.array(perm) - 1}
        \STATE img $\leftarrow$ \texttt{Image.open(image\_path)}
        \STATE img\_array $\leftarrow$ \texttt{np.array(img)}
        \STATE height, width, channels $\leftarrow$ \texttt{img\_array.shape}
        \STATE block\_size $\leftarrow$ \texttt{len(perm)}
        \FORALL{$i$ in \texttt{range(0, height, block\_size)}}
            \FORALL{$j$ in \texttt{range(0, width, block\_size)}}
                \STATE block $\leftarrow$ \texttt{img\_array[i:i + block\_size, j:j + block\_size]}
                \FORALL{$r$ in \texttt{range(block\_size)}}
                    \STATE block[$r, :$] $\leftarrow$ \texttt{block[$r, perm[r]$]}
                \ENDFOR
                \STATE img\_array[i:i + block\_size, j:j + block\_size] $\leftarrow$ block
            \ENDFOR
        \ENDFOR
        \STATE shuffled\_img $\leftarrow$ \texttt{Image.fromarray(img\_array.astype(np.uint8))}
        \STATE \texttt{shuffled\_img.save("shuffled\_image.png")}
    \end{algorithmic}
\end{algorithm}

\begin{algorithm}
    \caption{Rotate Image}
    \label{alg4}
    \begin{algorithmic}[1]
        \REQUIRE image\_path
        \ENSURE rotated image saved as "rotated\_image.png"
        \STATE img $\leftarrow$ \texttt{Image.open(image\_path)}
        \STATE rotated\_img $\leftarrow$ \texttt{img.rotate(-90)}
        \STATE \texttt{rotated\_img.save("rotated\_image.png")}
    \end{algorithmic}
\end{algorithm}

\begin{algorithm}
    \caption{Rotate Image Counter-Clockwise}
    \label{alg5}
    \begin{algorithmic}[1]
        \REQUIRE image\_path
        \ENSURE rotated image saved as "rotated\_image\_counter\_clockwise.png"
        \STATE img $\leftarrow$ \texttt{Image.open(image\_path)}
        \STATE rotated\_img $\leftarrow$ \texttt{img.rotate(90)}
        \STATE \texttt{rotated\_img.save("rotated\_image\_counter\_clockwise.png")}
    \end{algorithmic}
\end{algorithm}

\begin{algorithm}
    \caption{Unshuffle Pixels of Image}
    \label{alg6}
    \begin{algorithmic}[1]
        \REQUIRE image\_path
        \ENSURE unshuffled image saved as "unshuffled\_image.png"
        \STATE perm $\leftarrow$ \texttt{input from user or keys.txt}
        \STATE perm $\leftarrow$ \texttt{np.array(perm) - 1}
        \STATE img $\leftarrow$ \texttt{Image.open(image\_path)}
        \STATE img\_array $\leftarrow$ \texttt{np.array(img)}
        \STATE inv\_perm $\leftarrow$ \texttt{np.argsort(perm)}
        \STATE height, width, channels $\leftarrow$ \texttt{img\_array.shape}
        \STATE block\_size $\leftarrow$ \texttt{len(perm)}
        \FORALL{$i$ in \texttt{range(0, height, block\_size)}}
            \FORALL{$j$ in \texttt{range(0, width, block\_size)}}
                \STATE block $\leftarrow$ \texttt{img\_array[i:i + block\_size, j:j + block\_size]}
                \FORALL{$r$ in \texttt{range(block\_size)}}
                    \STATE block[$r, :$] $\leftarrow$ \texttt{block[$r, inv\_perm[r]$]}
                \ENDFOR
                \STATE img\_array[i:i + block\_size, j:j + block\_size] $\leftarrow$ block
            \ENDFOR
        \ENDFOR
        \STATE unshuffled\_img $\leftarrow$ \texttt{Image.fromarray(img\_array.astype(np.uint8))}
        \STATE \texttt{unshuffled\_img.save("unshuffled\_image.png")}
    \end{algorithmic}
\end{algorithm}

\begin{algorithm}
    \caption{Unshuffle and Unpad Image}
    \label{alg7}
    \begin{algorithmic}[1]
        \REQUIRE image\_path, original\_width, original\_height, SudokuSize
        \ENSURE unshuffled and unpadded image saved as "unshuffled\_and\_unpadded\_image.png"
        \STATE img $\leftarrow$ \texttt{Image.open(image\_path)}
        \STATE img\_array $\leftarrow$ \texttt{np.array(img)}
        \STATE width, height $\leftarrow$ \texttt{img.size}
        \STATE seed $\leftarrow$ \texttt{height}
        \STATE \texttt{np.random.seed(seed)}
        \STATE perm $\leftarrow$ \texttt{np.random.permutation(len(img\_array))}
        \STATE unshuffled\_img\_array $\leftarrow$ \texttt{np.empty\_like(img\_array)}
        \FOR{each $original\_index$, $shuffled\_index$ in \texttt{enumerate(perm)}}
            \STATE unshuffled\_img\_array[$shuffled\_index$] $\leftarrow$ img\_array[$original\_index$]
        \ENDFOR
        \STATE img $\leftarrow$ \texttt{Image.fromarray(unshuffled\_img\_array)}
        \STATE pad\_height $\leftarrow$ \texttt{SudokuSize - original\_height if original\_height < SudokuSize else (original\_height // SudokuSize + 1) * SudokuSize - original\_height if original\_height \% SudokuSize != 0 else 0}
        \STATE pad\_width $\leftarrow$ \texttt{SudokuSize - original\_width if original\_width < SudokuSize else (original\_width // SudokuSize + 1) * SudokuSize - original\_width if original\_width \% SudokuSize != 0 else 0}
        \STATE img $\leftarrow$ \texttt{img.crop((0, 0, img.width - pad\_width, img.height - pad\_height))}
        \STATE \texttt{img.save("unshuffled\_and\_unpadded\_image.png")}
    \end{algorithmic}
\end{algorithm}

\begin{algorithm}[H]
    \caption{Decrypt Threshold Image}
    \label{alg8}
    \begin{algorithmic}[1]
        \REQUIRE image\_path, randomNumber
        \ENSURE decrypted image saved as "decrypted\_image.png"
        \STATE img $\leftarrow$ \texttt{Image.open(image\_path)}
        \STATE img\_array $\leftarrow$ \texttt{np.array(img)}
        \IF{\texttt{len(img\_array.shape) == 3}} \COMMENT{Color image}
            \FOR{each $i$ in \texttt{range(img\_array.shape[0])}}
                \FOR{each $j$ in \texttt{range(img\_array.shape[1])}}
                    \FOR{each $k$ in \texttt{range(img\_array.shape[2])}}
                        \IF{img\_array[$i, j, k$] - randomNumber $>$ 0}
                            \STATE img\_array[$i, j, k$] $\leftarrow$ img\_array[$i, j, k$] - randomNumber
                        \ELSE
                            \STATE img\_array[$i, j, k$] $\leftarrow$ img\_array[$i, j, k$] + 255 - randomNumber
                        \ENDIF
                    \ENDFOR
                \ENDFOR
            \ENDFOR
        \ELSIF{\texttt{len(img\_array.shape) == 2}} \COMMENT{Grayscale image}
            \FOR{each $i$ in \texttt{range(img\_array.shape[0])}}
                \FOR{each $j$ in \texttt{range(img\_array.shape[1])}}
                    \IF{img\_array[$i, j$] - randomNumber $>$ 0}
                        \STATE img\_array[$i, j$] $\leftarrow$ img\_array[$i, j$] - randomNumber
                    \ELSE
                        \STATE img\_array[$i, j$] $\leftarrow$ img\_array[$i, j$] + 255 - randomNumber
                    \ENDIF
                \ENDFOR
            \ENDFOR
        \ENDIF
        \STATE decrypted\_img $\leftarrow$ \texttt{Image.fromarray(img\_array)}
        \STATE \texttt{decrypted\_img.save("decrypted\_image.png")}
    \end{algorithmic}
\end{algorithm}

\subsubsection{Algorithms for Videos}
This section presents an overview of the key algorithms utilized in video encryption.
\begin{algorithm}
    \caption{Video Encryption}
    \label{alg9}
    \begin{algorithmic}[1]
        \REQUIRE video\_path, output\_video\_name, sudoku\_perm\_file
        \ENSURE Processed video saved as output\_video\_name
        \STATE \texttt{image\_folder} $\leftarrow$ \texttt{'vid'}
        \STATE \texttt{video} $\leftarrow$ \texttt{cv2.VideoCapture(video\_path)}
        \STATE \texttt{perm} $\leftarrow$
        \WHILE {\texttt{video.isOpened()}}
            \STATE \texttt{ret, frame} $\leftarrow$ \texttt{video.read()}
            \IF {\texttt{ret}}
                \STATE \texttt{Convert the frame to image}
                \STATE \texttt{image\_path} $\leftarrow$ \texttt{f"\{image\_folder\}/frame\_\{cnt\}.png"}
                \STATE \texttt{Save the image at (image\_path)}
                \STATE \texttt{Apply image thresholding}
                \STATE \texttt{Apply padding and shuffling to the frame}
                \STATE \texttt{Apply Sudoku based transformations to encrypt the image}
                \STATE \texttt{Rotate the encrypted image}        
                \STATE \texttt{Save the image as (f"\{image\_folder\}/frame\_enc\_\{cnt\}.png")}
            \ELSE
                \STATE \texttt{break}
            \ENDIF
        \ENDWHILE
        \STATE \texttt{images} $\leftarrow$ \texttt{[img for img in os.listdir(image\_folder) if img.startswith("frame\_enc")]}
        \STATE \texttt{images.sort(key=lambda x: int(x.split('\_')[2].split('.')[0]))}
        \STATE \texttt{first\_frame} $\leftarrow$ \texttt{cv2.imread(os.path.join(image\_folder, images[0]))}
        \STATE \texttt{height, width, layers} $\leftarrow$ \texttt{first\_frame.shape}
        \STATE \texttt{video\_writer} $\leftarrow$ \texttt{cv2.VideoWriter(output\_video\_name, fourcc, fps, (width, height))}
        \FORALL {\texttt{image} in \texttt{images}}
            \STATE \texttt{video\_writer.write(cv2.imread(os.path.join(image\_folder, image)))}
        \ENDFOR
    \end{algorithmic}
\end{algorithm}

\subsubsection{Algorithms for Audios}
This section presents an overview of the key algorithms utilized in image processing.

\begin{algorithm}
    \caption{Audio Sudoku Based Shuffling}
    \label{alg10}
    \begin{algorithmic}[1]
        \REQUIRE audio\_path, perm
        \ENSURE Shuffled audio file saved as \texttt{'shuffled\_audio.wav'}
        \STATE \texttt{sample\_rate, audio\_data} $\leftarrow$ \texttt{wavfile.read(audio\_path)}
        \STATE \texttt{block\_size} $\leftarrow$ \texttt{len(perm)}
        \STATE \texttt{num\_blocks} $\leftarrow$ \texttt{len(audio\_data) // block\_size}
        \STATE \texttt{shuffled\_audio} $\leftarrow$ \texttt{np.zeros\_like(audio\_data)}
        
        \WHILE {i < num\_blocks}
            \STATE \texttt{start\_idx} $\leftarrow$ \texttt{i * block\_size}
            \STATE \texttt{end\_idx} $\leftarrow$ \texttt{start\_idx + block\_size}
            \STATE \texttt{shuffled\_block} $\leftarrow$ \texttt{audio\_data[start\_idx:end\_idx][perm]}
            \STATE \texttt{shuffled\_audio[start\_idx:end\_idx]} $\leftarrow$ \texttt{shuffled\_block}
        \ENDWHILE
        
        \IF {len(audio\_data) \% block\_size != 0}
            \STATE \texttt{remaining\_samples\_start\_idx} $\leftarrow$ \texttt{num\_blocks * block\_size}
            \STATE \texttt{shuffled\_audio[remaining\_samples\_start\_idx:]} $\leftarrow$ \texttt{audio\_data[remaining\_samples\_start\_idx:]}
        \ENDIF
        
        \STATE \texttt{wavfile.write('shuffled\_audio.wav', sample\_rate, shuffled\_audio)}
    \end{algorithmic}
\end{algorithm}

\begin{algorithm}
    \caption{Unshuffle Audio Blocks}
    \label{alg11}
    \begin{algorithmic}[1]
        \REQUIRE audio\_path, perm
        \ENSURE Unshuffled audio file saved as \texttt{'unshuffled\_audio.wav'}
        \STATE \texttt{sample\_rate, audio\_data} $\leftarrow$ \texttt{wavfile.read(audio\_path)}
        \STATE \texttt{block\_size} $\leftarrow$ \texttt{len(perm)}
        \STATE \texttt{num\_blocks} $\leftarrow$ \texttt{len(audio\_data) // block\_size}
        \STATE \texttt{shuffled\_audio} $\leftarrow$ \texttt{np.zeros\_like(audio\_data)}
        \STATE \texttt{inv\_perm} $\leftarrow$ \texttt{np.argsort(perm)}
        
        \WHILE{i < num\_blocks}
            \STATE \texttt{start\_idx} $\leftarrow$ \texttt{i * block\_size}
            \STATE \texttt{end\_idx} $\leftarrow$ \texttt{start\_idx + block\_size}
            \STATE \texttt{shuffled\_block} $\leftarrow$ \texttt{audio\_data[start\_idx:end\_idx][inv\_perm]}
            \STATE \texttt{shuffled\_audio[start\_idx:end\_idx]} $\leftarrow$ \texttt{shuffled\_block}
        \ENDWHILE
        
        \IF {len(audio\_data) \% block\_size != 0}
            \STATE \texttt{remaining\_samples\_start\_idx} $\leftarrow$ \texttt{num\_blocks * block\_size}
            \STATE \texttt{shuffled\_audio[remaining\_samples\_start\_idx:]} $\leftarrow$ \texttt{audio\_data[remaining\_samples\_start\_idx:]}
        \ENDIF
        
        \STATE \texttt{wavfile.write('unshuffled\_audio.wav', sample\_rate, shuffled\_audio)}
    \end{algorithmic}
\end{algorithm}

\begin{algorithm}
    \caption{Audio Sudoku Based XOR Encryption}
    \label{alg12}
    \begin{algorithmic}[1]
        \REQUIRE sudoku\_board
        \ENSURE Encrypted audio file saved as \texttt{'encrypted\_audio.wav'}
        \STATE \texttt{sample\_rate, audio\_data} $\leftarrow$ \texttt{wavfile.read('filename')}
        \STATE \texttt{block\_size} $\leftarrow$ 9 \COMMENT{For a 9x9 Sudoku}
        \STATE \texttt{padded\_length} $\leftarrow$ \texttt{int(np.ceil(len(audio\_data) / block\_size) * block\_size)}
        \STATE \texttt{padded\_audio} $\leftarrow$ \texttt{np.pad(audio\_data, (0, padded\_length - len(audio\_data)), 'constant')}
        \STATE \texttt{blocks} $\leftarrow$ \texttt{padded\_audio.reshape(-1, block\_size)}
        
        \WHILE {i < blocks.shape[0]}
            \WHILE {j < blocks.shape[1]}
                \STATE \texttt{blocks[i, j]} $\leftarrow$ \texttt{np.bitwise\_xor(blocks[i, j], sudoku\_board[i \% 9][j \% 9])}
            \ENDWHILE
        \ENDWHILE
        
        \STATE \texttt{blocks} $\leftarrow$ \texttt{np.transpose(blocks)}
        \STATE \texttt{encrypted\_audio} $\leftarrow$ \texttt{blocks.flatten()}
        \STATE \texttt{encrypted\_audio} $\leftarrow$ 
        \STATE \texttt{wavfile.write('encrypted\_audio.wav', sample\_rate, encrypted\_audio)}
    \end{algorithmic}
\end{algorithm}

\subsection{Steps for Image}

\subsubsection{Encryption Steps for Image}

To convert the input data into a highly secure encrypted format, there are multiple steps in the encryption process. The encryption process goes as follows:

\begin{itemize}

    \item \textbf{Thresholding:} As shown in algorithm \ref{alg1}, the input file undergoes initial conversion into a binary format, followed by the addition of a random threshold value to the bits. By incorporating a random threshold value to the binary bits, an additional layer of variability and complexity is introduced, enhancing the security measures of the encryption process
    
    \item \textbf{Padding and Shuffling:} The file is padded to a size equal to a multiple of the Sudoku grid size, ensuring compatibility with the Sudoku-based transformations. The blocks are then shuffled according to a pattern derived from a Sudoku puzzle, enhancing security. Algorithm \ref{alg2} shows the steps taken.
    
    \item \textbf{Sudoku Transformation:} Following that, the shuffled media undergoes a transformation based on the solution of a Sudoku puzzle as depicted in algorithm \ref{alg3}. This transformation involves shuffling the bits in rows, grouped in blocks of size of the Sudoku, and shuffling the rows in batches of size of the Sudoku. This process leads to a complex and non-linear rearrangement of the data, enhancing the algorithm's robustness.

    \item \textbf{Rotation:} Another layer of security can be added to the algorithm by changing the orientation by, rotating the image (or transposing the audio) 90 degrees clockwise which is done in algorithm \ref{alg4}.
    
\end{itemize}

\subsubsection{Decryption Steps for Image} The decryption process reverses the encryption steps to recover the original data. The steps are as follows:

\begin{itemize}
    \item \textbf{Reverse Rotation:} The decrypted image is rotated 90 degrees counter-clockwise, or the audio is transposed again as algorithm \ref{alg5} shows.
    \item \textbf{Reverse Sudoku Transformation:} Applies the inverse Sudoku puzzle transformations to the rotated image, steps for which are shown in \ref{alg6}.
    \item \textbf{Unshuffling and Unpadding:} Reverts the shuffling and removes padding to restore the original image size as shown in \ref{alg7}.
    \item \textbf{Decrypt Thresholding:} Algorithm \ref{alg8} converts the binary image back to its original format by reversing the thresholding process, by subtracting the threshold value.
\end{itemize}

\subsection{Steps for Video}
Encrypting videos using this algorithm works similarly to the encryption of images. Videos are a sequence of images known as frames so encrypting each frame with the proposed algorithm and then combining them with the same frame rate as the original video gives an encrypted version of the video ~\hyperref[ref8]{[8]}. This can be improved by checking for changes in the frame and the next frame and only encrypting them when a change has been detected, thereby reducing the number of images passed to the algorithm.

Videos can be encrypted in two primary ways: secret key encryption and asymmetric encryption. Secret key encryption, a part of cryptographic algorithms, uses a single key for both encoding and decoding processes ~\hyperref[ref9]{[9]}. In asymmetric encryption, both private and public keys are used to encrypt and decrypt data. There are various approaches to video encryption. Selective encryption encrypts only a subset of the data, meaning not every byte of the video stream is encrypted; instead, only selected frames are encrypted. This method results in lower security but faster processing times. Full encryption involves compressing the entire video and then encrypting it using traditional algorithms like AES and RSA. While this method offers higher security, it is computationally intensive and not suitable for real-time applications.

This work's approach as shown in algorithm \ref{alg9} employs a symmetric key and operates on each frame of the video individually, executing encryption as if each frame were an image, performing the steps of Thresholding, Padding, Shuffling, Sudoku-based Encryption and Rotation. Subsequently, the encrypted frames are combined to form the resulting video. While this method may be slower, it significantly enhances security by ensuring thorough encryption at the frame level, thereby safeguarding the entire video content from unauthorized access or tampering.

\subsection{Steps for Audio}
Audio Encryption can be done using both shuffling the bits (refer algorithms \ref{alg10} and \ref{alg11}) and XOR operations that are added to a Sudoku-based permutation technique. The steps include reading the audio file, reshaping it into blocks, and adding padding to ensure its length is multiple of the block size (which is the size of the Sudoku grid). 
Methods mentioned previously in image and video encryption can also be modified to be used for audio ~\hyperref[ref12]{[12]}, which is simpler since audio can be represented in 1-dimensional blocks on which sudoku transformations can be applied.
\\
Alternatively, XOR operation is used to encrypt these blocks which are depicted in algorithm \ref{alg12}. This step in the encryption process involves applying an XOR operation between the transformed image and a key derived from the Sudoku puzzle. This substitution method ensures that each pixel value is altered in a non-predictable manner, providing strong encryption.
The audio blocks are then shuffled using a function which generates a shuffled audio file by using a permutation array derived from the Sudoku values. After loading the original and shuffled audio signals, the code plots the waveforms and extracts characteristics such as the spectral centroid, spectral bandwidth, zero-crossing rate, and RMS energy. To determine the effect of the encryption, these features are printed and contrasted. A function in the code also computes the Signal-to-Noise Ratio (SNR) between the original and shuffled audio, offering a numerical representation of the quality reduction brought on by encryption. This thorough procedure evaluates the impact of Sudoku-based encryption on audio quality and shows how it can be used with audio data.

\section{Experimental Results}
The proposed algorithm is assessed using various forms of multimedia like images, videos and audio in various iterations to improve the quality of the sample. 

\subsection{Image Results}
\begin{figure}
  \centering
  \begin{minipage}{.32\textwidth}
    \centering
    \includegraphics[width=\linewidth]{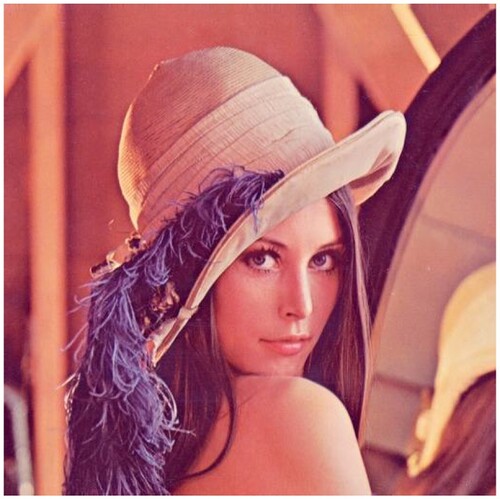}
    \captionof{figure}{Original Image before encryption}
    \label{fig:original_image}
  \end{minipage}%
  \hfill
  \begin{minipage}{.32\textwidth}
    \centering
    \includegraphics[width=\linewidth]{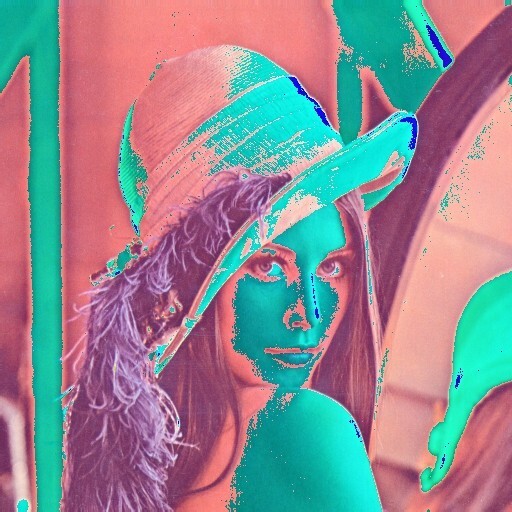}
    \captionof{figure}{Image after Thresholding}
    \label{fig:enc_thresholding}
  \end{minipage}%
  \hfill
  \begin{minipage}{.32\textwidth}
    \centering
    \includegraphics[width=\linewidth]{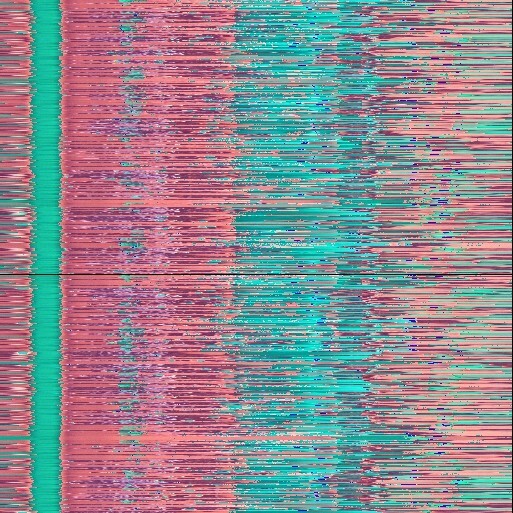}
    \captionof{figure}{Image after Padding and Shuffling}
    \label{fig:enc_padding_shuffling}
  \end{minipage}

  \begin{minipage}{.32\textwidth}
    \centering
    \includegraphics[width=\linewidth]{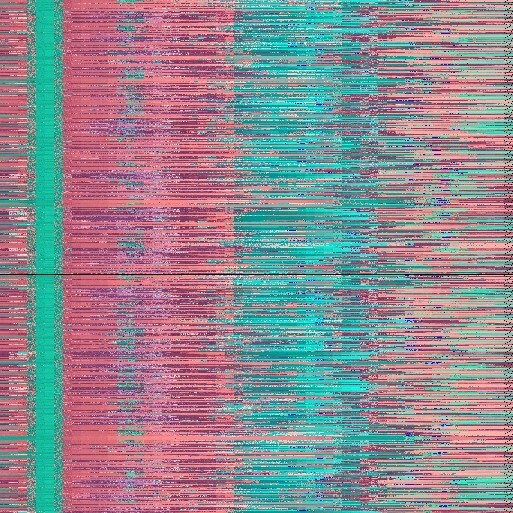}
    \captionof{figure}{Image after applying Sudoku key}
    \label{fig:enc_sudoku_key}
  \end{minipage}%
  \hfill
  \begin{minipage}{.32\textwidth}
    \centering
    \includegraphics[width=\linewidth]{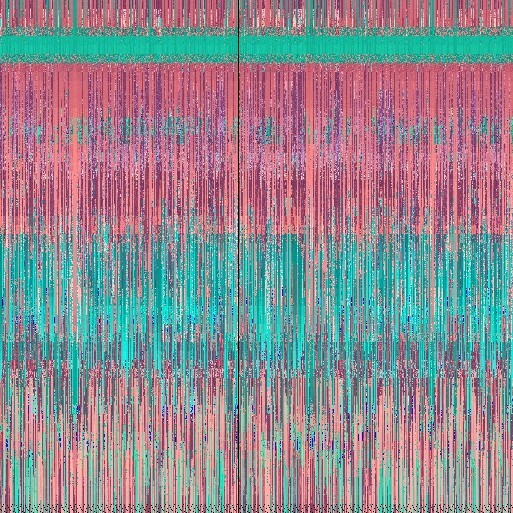}
    \captionof{figure}{Image after rotation}
    \label{fig:enc_rotation}
  \end{minipage}%
  \hfill
  \begin{minipage}{.32\textwidth}
    \centering
    \includegraphics[width=\linewidth]{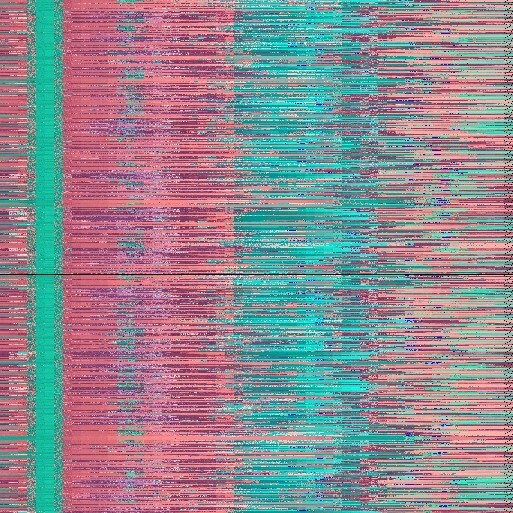}
    \captionof{figure}{Image after Rotating Anticlockwise}
    \label{fig:rotating_anticlockwise}
  \end{minipage}

  \begin{minipage}{.32\textwidth}
    \centering
    \includegraphics[width=\linewidth]{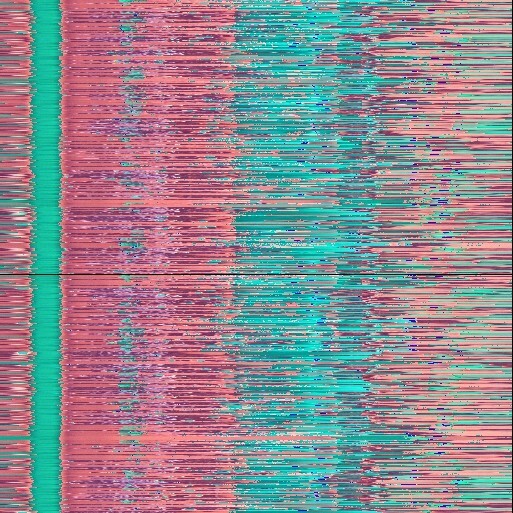}
    \captionof{figure}{Image after Decrypting Sudoku Key}
    \label{fig:dec_sudoku_key}
  \end{minipage}%
  \hfill
  \begin{minipage}{.32\textwidth}
    \centering
    \includegraphics[width=\linewidth]{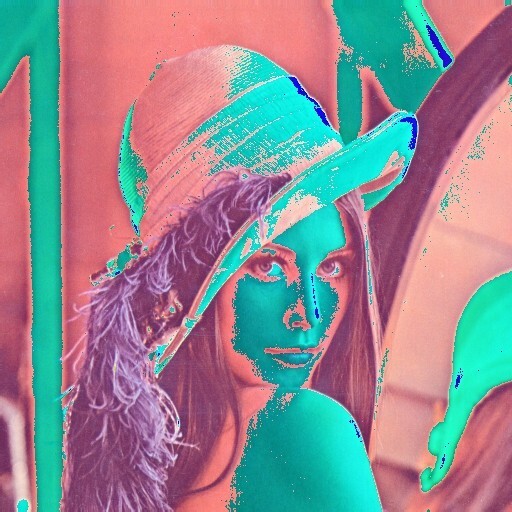}
    \captionof{figure}{Image after unshuffling \& unpadding}
    \label{fig:dec_unpadding_unshuffling}
  \end{minipage}%
  \hfill
  \begin{minipage}{.32\textwidth}
    \centering
    \includegraphics[width=\linewidth]{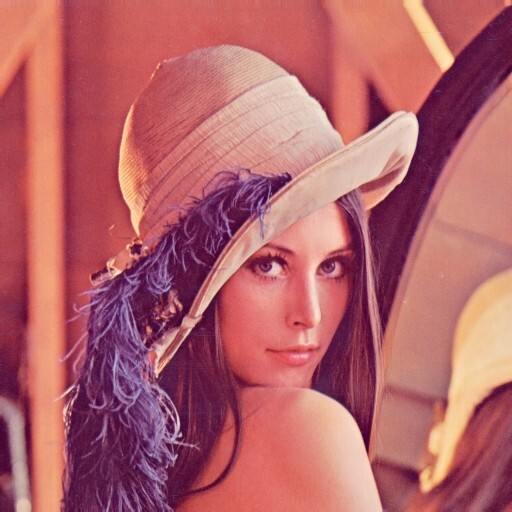}
    \captionof{figure}{Image after Decryption}
    \label{fig:dec_image}
  \end{minipage}

\end{figure}
% \subsubsection{Encryption and Decryption Process}
As shown in the \emph{Fig}. \emph{\ref{fig:original_image}}, the image of \textquotesingle Lena\textquotesingle{} ~\hyperref[ref13]{[13]}  was used for the encryption and decryption. The size of the image was 512x512. The thresholding of images was observed for various iterations. The padding and shuffling make the image unnoticeable to an attacker. The key generation using 9x9 Sudoku and 16x16 Sudoku has a user input setup where the user sets up the key for encryption. Finally, the image is rotated again to make it computationally stronger to decrypt by an attacker. The encryption and decryption process has the following steps. \emph{Fig}. \emph{\ref{fig:original_image}}  chooses an original image. \emph{Fig}. \emph{\ref{fig:enc_thresholding}} adds thresholding to the pixels of the image converting the data to a binary format, \emph{Fig}. \emph{\ref{fig:enc_padding_shuffling}} adds padding so that Sudoku transformations can be applied to it, shuffling it based on a Sudoku puzzle, applying the Sudoku transformation, and finally \emph{Fig}. \emph{\ref{fig:enc_rotation}} rotates the image for further security. The encryption key is generated dynamically using the current timestamp, ensuring each session has a unique key. 
The first step in the decryption process is to reverse the rotation applied during encryption. The image is rotated back to its original orientation as shown in \emph{Fig}. \emph{\ref{fig:rotating_anticlockwise}}. The image, after being unrotated, undergoes a reverse Sudoku transformation. This step involves undoing the shuffling and transformation based on the Sudoku puzzle used during encryption resulting in \emph{Fig}. \emph{\ref{fig:dec_sudoku_key}}. \emph{Fig}. \emph{\ref{fig:dec_unpadding_unshuffling}} shows the padded portions of the image removed to restore it to its original dimensions. Finally, the system outputs the encrypted or decrypted files, ready for secure storage or transmission giving the final \emph{Fig}. \emph{\ref{fig:dec_image}}. This approach provides strong security and efficient processing for different types of media.
\subsection{Video Results}
\begin{figure}
    \centering
    \includegraphics[width=0.8\linewidth]{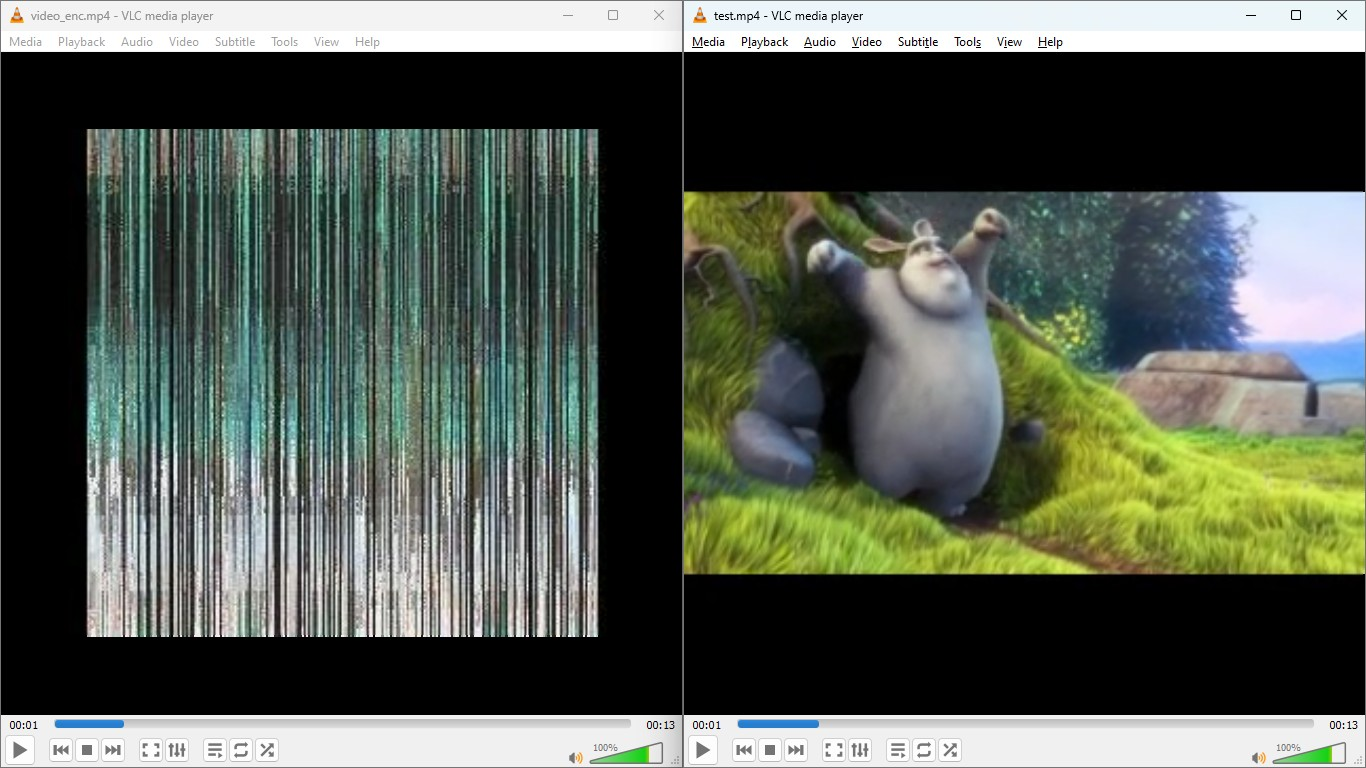}
    \caption{Comparison of encrypted and original video}
    \label{fig:video_result}
\end{figure}

As seen in \emph{Fig}. \emph{\ref{fig:video_result}}, the video encryption process follows a methodology similar to that of image encryption. In this approach, the video is segmented into individual frames, and each frame is encrypted sequentially. By encrypting each frame one by one, the entire video is transformed into its encrypted form. This technique ensures that the video data is securely protected frame by frame, maintaining the integrity and confidentiality of the content throughout the encryption process.

General cryptographic algorithms are not usually suited for encrypting video data as they cannot process the large volume of video data in real-time ~\hyperref[ref14]{[14]}, this Sudoku based approach can provide a base for future algorithms that can use the security provided by this algorithm and also work real-time.

\subsection{Audio Results}
\begin{figure}[H]
    \centering
    \begin{minipage}{0.32\linewidth}
        \centering
        \includegraphics[width=\linewidth]{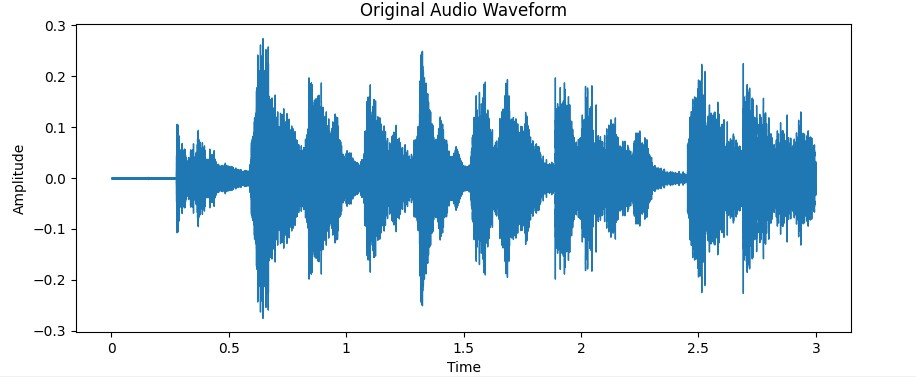}
        \caption{Waveform of original audio}
        \label{fig:og_audio}
    \end{minipage}
    \hfill
    \begin{minipage}{0.32\linewidth}
        \centering
        \includegraphics[width=\linewidth]{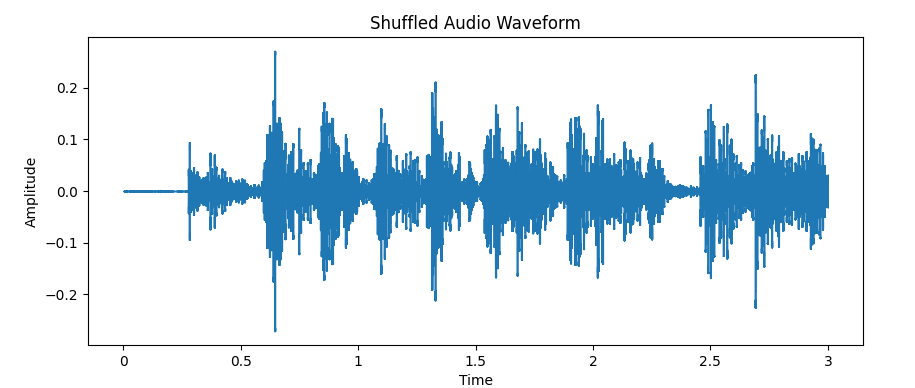}
        \caption{Shuffled Audio Waveform}
        \label{fig:shuf_audio}
    \end{minipage}
    \hfill
    \begin{minipage}{0.32\linewidth}
        \centering
        \includegraphics[width=\linewidth]{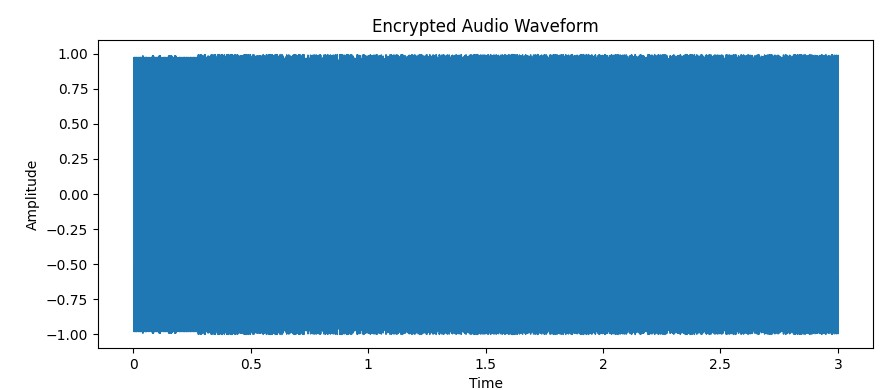}
        \caption{XOR Audio Waveform}
        \label{fig:xor_audio}
    \end{minipage}
\end{figure}

\emph{Fig}. \emph{\ref{fig:og_audio}}, \emph{Fig}. \emph{\ref{fig:shuf_audio}}, \emph{Fig}. \emph{\ref{fig:xor_audio}} shows the waveforms of the original, shuffled, and substituted audios respectively, which provide a clear visualization of how the audio changes after encryption. It is evident that the more the waveforms differ, the more secure and challenging it becomes for an attacker to decrypt without the necessary keys. The paper compares the original audio waveform to the audio waveforms resulting from Sudoku-based XOR transformations and Sudoku-based shuffling. The significant differences observed in the shuffled and XOR-based waveforms from the original audio demonstrate the effectiveness of these encryption methods in enhancing security and making decryption difficult.

% \subsection{Applications}

\section{Analysis}
The proposed algorithm is examined using a variety of metrics for different types of media. Initially, key analysis investigates the possible pairings of the suggested method to minimize key size for transmission or storage. Second, image analysis evaluates how the algorithm works on different images using Number of Pixels Change Rate (NPCR) and Unified Average Changing Intensity (UACI) values. Third, the audio analysis compares Signal-to-noise ratio (SNR),  Peak signal-to-noise ratio (PSNR) and Mean Squared Error (MSE) values.
\subsection{Key Analysis}
The key space of an encryption algorithm is fundamental in evaluating its resistance to brute-force attacks. A larger key space implies more possible keys, which exponentially increases the difficulty of successfully guessing the correct key.

1. Sudoku Grid Combinations
The user puts in the grid combinations for the Sudoku-based encryption method, the key space is primarily influenced by the number of solvable Sudoku grids, which is approximately 6.67 sextillion making it extremely easy for the user to feed data.

2. Key Space and Brute Force Resistance:
Because of the large key space, brute-forcing is not feasible. For example, if all the possible combinations for a 400 × 400 image with 100 encryption rounds on the fastest supercomputer, the time would be:
% 42.21 × 1 0 892 combinations ÷ 1.714\% × 1 0 18 operations/second ≈ 7.76 × 1 0 881 years
% 4.2 x 10\^9 combinations = 1.714 x 10^18 operations/second =
$7.76 \times 10^{881}$ years
. This long duration ensures robust security by demonstrating the impracticality of brute-force attacks.

3. NP-Complete Nature and Timestamp-based Keys:
Sudoku is also an NP-complete problem and thus, cannot be solved in polynomial time. In addition to the vast number of Sudoku solutions, incorporating timestamp-based key generation adds another layer of complexity. Timestamps introduce a temporal element that ensures keys are unique over time, making it even more challenging for an attacker to predict or replicate the key. Furthermore, encrypting the Sudoku key with robust cryptographic algorithms like ECC, RSA ~\hyperref[ref15]{[15]}, or AES ~\hyperref[ref16]{[16]} adds another security layer. 

When these elements are combined many Sudoku solutions, dynamic timestamp based keys, and extra common cryptographic algorithms a very big and intricate key space is produced. This ensures strong security for encrypted images by making the Sudoku-based encryption technique extremely resistant to brute-force attacks.

\subsubsection{Key Generation time}
Regarding the running time assessments, a Windows 11 laptop computer with 8 gigabytes of RAM, and running on a processor with the conﬁgurations: Intel(R) Core (TM) i5-1135G7 CPU @ 2.40GHz base speed with Turbo Boost up to 4.20GHz was used. The proposed algorithm was tested on several iterations.
Py-Sudoku is a Python package designed to facilitate working with Sudoku puzzles. It typically includes features for generating, solving, and validating Sudoku grids.

% \begin{table}[h]
%     \centering
%     \caption{Key Generation Time}
%     \renewcommand{\arraystretch}{1.3} % Increase the row height by 50%
%     \begin{tabular}{@{}lcccc@{}}
%         \toprule
%         \textbf{Number of Keys (9x9 Sudoku)} & \hspace{2cm} \textbf{Time Required (seconds)} \\ \midrule
%         10 & 0.022224903106689453 \\
%         25 & 0.08881902694702148 \\
%         50 & 0.2174370288848877 \\
%         75 & 0.39699816703796387 \\
%         100 & 0.639775276184082 \\ \bottomrule
%     \end{tabular}
%     \label{tab:TimeComparison}
% \end{table}

\begin{table}[H]
    \centering
    \caption{Key Generation Time}
    \renewcommand{\arraystretch}{1.3} % Increase the row height for readability
    \setlength{\tabcolsep}{10pt} % Increase column spacing
    \begin{tabular}{lcc}
        \toprule
        \textbf{Number of Keys (9x9 Sudoku)} & \textbf{Time Required (s)} \\
        \midrule
        10 & 0.022 \\
        25 & 0.089 \\
        50 & 0.217 \\
        75 & 0.397 \\
        100 & 0.640 \\
        \bottomrule
    \end{tabular}
    \label{tab:TimeComparison}
\end{table}

\emph{Table} \emph{\ref{tab:TimeComparison}} shows the time taken to generate a 9x9 Sudoku grid over multiple iterations (25, 50, 75, 100). It demonstrates that as the number of iterations increases, the generation time also increases, thus it also makes the decryption more difficult with a higher number of keys.

\subsubsection{Alternative approach}
The Sudoku algorithm begins with an empty n x n board, filling diagonal sub-grids with unique random numbers from 1 to n. It then uses backtracking to solve, trying numbers in empty cells, checking validity, and backtracking if needed until a solution is found. It's adaptable to other characters/alphabets and verifies correctness, detecting tampering if the Sudoku is incorrect.

% \begin{table}[h]
%     \centering
%     \caption{Comparison of generation of Sudoku}
%     \renewcommand{\arraystretch}{1.5} % Increase the row height by 50%
%     \begin{tabular}{@{}lcccc@{}}
%         \toprule
%         \textbf{Size of Sudoku} & \hspace{2cm} \textbf{Time Required (s)} \\ \midrule
%         4x4 & 0.0002 \\
%         9x9 & 0.0053 \\
%         16x16 & 0.0985 \\
%         25x25 & 0.7554 \\ \bottomrule
%     \end{tabular}
%     \label{tab:sudoku_size_time}
% \end{table}

\begin{table}[H]
    \centering
    \caption{Comparison of Sudoku Generation Times}
    \renewcommand{\arraystretch}{1.3} % Increase row height for readability
    \setlength{\tabcolsep}{12pt} % Increase column spacing
    \begin{tabular}{lcc}
        \toprule
        \textbf{Size of Sudoku} & \textbf{Time Required (s)} \\
        \midrule
        4x4 & 0.0002 \\
        9x9 & 0.0053 \\
        16x16 & 0.0985 \\
        25x25 & 0.7554 \\
        \bottomrule
    \end{tabular}
    \label{tab:sudoku_size_time}
\end{table}

As shown in the \emph{table} \emph{\ref{tab:sudoku_size_time}}, a 4x4 Sudoku takes the shortest time but has security concerns as it is extremely easy to decrypt. For that matter, 9x9 and 16x16 give good results concerning the security of the encryption key as well.

\subsubsection{Security Analysis}
Sudoku puzzles inherently offer significant security due to their NP-complete nature. Solving a Sudoku puzzle is computationally intensive, especially for larger grids, making it difficult to reverse-engineer the key. The key size for a Sudoku-based encryption method can be extremely large, dependent on the number of valid Sudoku grids, which provides immense space for cryptographic keys. Moreover, the use of characters instead of numbers in the Sudoku grid can enhance security by increasing the complexity and variability of the puzzle, adding another layer of difficulty for potential attackers.
Furthermore, using unsolved Sudoku puzzles as keys can provide an additional security advantage. Since solving a Sudoku is NP-complete, transmitting unsolved Sudoku puzzles as keys complicates the attacker's task. There exist various algorithms like Backtracking, brute force, exact cover etc which can solve those Sudokus, but take more time. This approach ensures that even if the Sudoku puzzle is intercepted, the encrypted clues and their positions remain secure, making it nearly impossible for attackers to crack the encryption without access to the original clues and their positions.
Encrypting with robust cryptographic algorithms like ECC, RSA, and AES ~\hyperref[ref17]{[17]} can significantly enhance the security of the overall scheme. 

% \begin{figure}
%     \centering
%     \includegraphics[width=0.5\linewidth]{sudocomparision.png}
%     \caption{Enter Caption}
%     \label{fig:enter-label}
% \end{figure}
% There also exists a novel encryption algorithm called SudoKrypt.

\subsection{Image Analysis}
This section describes the analysis done of different forms of image data.
\subsubsection{Time Taken for each Step}
The \emph{Fig}. \emph{\ref{fig:img_time_steps}} below shows the time taken for each step of the algorithm used for a single iteration of the image used for encryption and decryption. From the graph, it is evident that the thresholding step consumes majority of the time. In this step, a random value can be added to each pixel of the image, resulting in a high time complexity of O(N
\textsuperscript{3})

\begin{figure}
    \centering
    \includegraphics[width=1\linewidth]{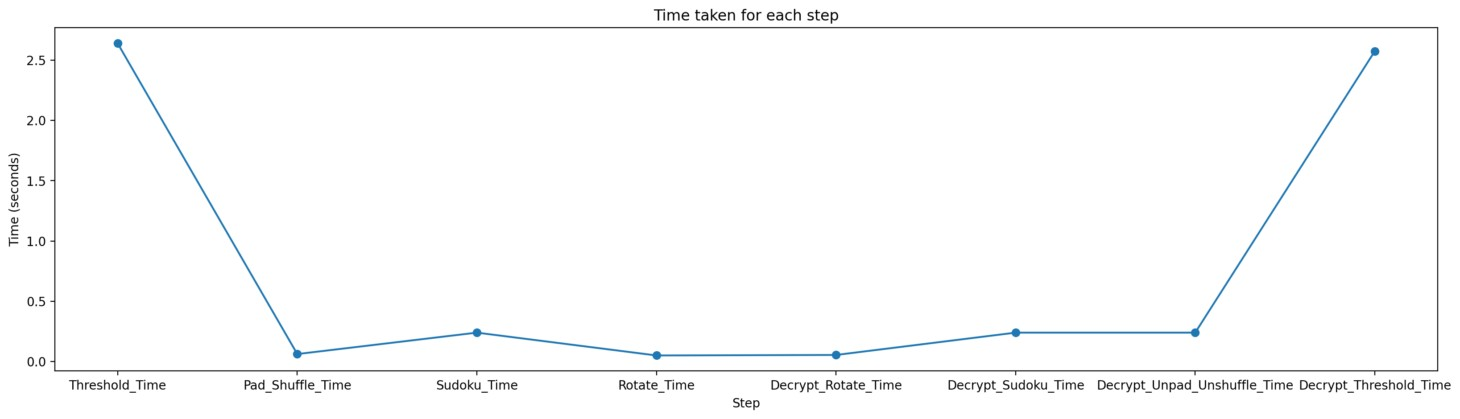}
    \caption{Time taken for Each Step}
    \label{fig:img_time_steps}
\end{figure}

\subsubsection{Time Taken for Iterations}
\emph{Table}. \emph{\ref{tab:sudoku_size_itr_time}} shows the time taken for different numbers of iterations when applying the Sudoku algorithm. It can be seen that the algorithm completes 100 iterations in just 12 seconds, highlighting its computational efficiency. This analysis excludes the time needed for key generation and only focuses on the encryption process.

\begin{table}[H]
    \centering
    \caption{Comparison of Time and Number of Iterations}
    \renewcommand{\arraystretch}{1.3} % Increase the row height for readability
    \setlength{\tabcolsep}{10pt} % Increase column spacing
    \begin{tabular}{lcc}
        \toprule
        \textbf{Number of Iterations} & \textbf{Time Required (s)} \\
        \midrule
        25 & 3.59 \\
        50 & 6.15 \\
        75 & 9.74 \\
        100 & 12.39 \\
        \bottomrule
    \end{tabular}
    \label{tab:sudoku_size_itr_time}
\end{table}

% \subsubsection{Histogram Analysis}
% The histogram of pixel values in an image serves as a representation of its color or intensity distribution. A robust encryption method should result in encrypted images with histograms significantly different from those of the original images. This dissimilarity reduces the susceptibility of the encrypted images to statistical attacks.

% \begin{figure}[H]
%     \centering
%     \includegraphics[width=0.5\linewidth]{tower_histogram.png}
%     \caption{Histogram for tower.png}
%     \label{fig:enter-label}
% \end{figure}
% \begin{figure}[H]
%     \centering
%     \includegraphics[width=0.5\linewidth]{tower_enc_histogram.png}
%     \caption{Histogram for encrypted image}
%     \label{fig:enter-label}
% \end{figure}

\subsubsection{Comparison with Other Images}
This section compares various images used for encryption. The time required for encryption having 100 iterations and using 9x9 Sudokus as the key is shown in \emph{Table}. \emph{\ref{tab:total_time_comparison}}. The time is equal to the time required for thresholding, padding, applying the Sudoku-based algorithm for shuffling the blocks, and rotation.

% \begin{table}[htbp]
%     \centering
%     \caption{Total time taken for 100 iterations across all images}
%     \begin{tabular}{llll}
%         \hline
%         \textbf{Image} & \textbf{Resolution} & \textbf{Sudoku Used} & \textbf{Time Taken (s)} \\
%         \hline
%         {Camerman} & {256x256} & 9x9 & 6.387107610702515 \\
%         {Lena} & {512x512} & 9x9 & 23.696540594100952 \\
%         {House} & {512x512} & 9x9 & 40.81598520278931 \\
%         {Mandarill} & {512x512} & 9x9 & 37.12921595573425 \\
%         {San Diego} & {1024x1024} & 9x9 & 147.18497920036316 \\
%         {Towers} & {800x1210} & 9x9 & 63.343812465667725 \\
%         \hline
%     \end{tabular}
%     \label{tab:total_time_comparison}
% \end{table}

\begin{table}[htbp]
    \centering
    \caption{Total Time Taken for 100 Iterations Across All Images}
    \renewcommand{\arraystretch}{1.3} % Increase row height for readability
    \setlength{\tabcolsep}{12pt} % Increase column spacing
    \begin{tabular}{lcccc}
        \toprule
        \textbf{Image} & \textbf{Resolution} & \textbf{Sudoku Used} & \textbf{Time Taken (s)} \\
        \midrule
        Camerman & 256x256 & 9x9 & 6.39 \\
        Lena & 512x512 & 9x9 & 23.70 \\
        House & 512x512 & 9x9 & 40.82 \\
        Mandarill & 512x512 & 9x9 & 37.13 \\
        San Diego & 1024x1024 & 9x9 & 147.18 \\
        Towers & 800x1210 & 9x9 & 63.34 \\
        \bottomrule
    \end{tabular}
    \label{tab:total_time_comparison}
\end{table}

Thus from the \emph{table} \emph{\ref{tab:total_time_comparison}}, it can analyzed that the algorithm works on all types of images from grayscale to color, from all sizes including squares (256x256 and its multiples), sizes not divisible by the sudoku size like 512x512 are not divisible by a 9x9 size thus padding is added to increase its size to 513x513, other sizes like 800x1024 and so on. The time increases as usage of non-multiples of Sudoku size and increase the image resolution. Different sizes of sudokus like 9x9, 16x16, and so on, which also affects the time taken.

\emph{table} \emph{\ref{tab:image_pixel_intensities}} presents a comparison of average pixel intensities between original images and encrypted images using 9x9 Sudoku-based encryption. The noticeable change in average pixel values post-encryption underscores the effectiveness of the encryption in introducing confusion and complexity, making it difficult for attackers to deduce information from the encrypted images through statistical analysis.

% \begin{tabular}{@{}lccccccl@{}}
% \hline Image& \multicolumn{3}{c}{ Original Image Pixel Intensities } & \multicolumn{3}{c}{ Encrypted Image Pixel Intensities } \\
% & Red & Green & Blue & Red & Green & Blue \\
% \hline Lena& 180.2237& 99.0512& 105.4103& 142.7590& 137.4564& 144.8255\\
% House & 155.4357 & 168.2260 & 142.2094 & 174.6491 & 184.1435 & 176.9407 \\
% Mandarill& 137.3913 & 128.8588 & 113.1171 & 147.4320 & 167.5345 & 124.8967 \\
% San Diego& 118.1405 & 175.6401 & 205.7076 & 157.5214 & 212.8628 & 239.9281 \\
% Towers& 97.0915& 127.5518& 155.8320& 83.3112 & 100.4656 & 96.2014
% \\\hline
% \end{tabular}

\begin{table}[H]
\centering
\caption{Comparison of original and encrypted image pixel intensities across different images.}
\renewcommand{\arraystretch}{1.3} % Increase row height for readability
\begin{tabular}{@{}lccccccl@{}}
\hline
Image & \multicolumn{3}{c}{Original Image Pixel Intensities} & \multicolumn{3}{c}{Encrypted Image Pixel Intensities} \\
 & Red & Green & Blue & Red & Green & Blue \\
\hline
Lena & 180.23 & 99.05 & 105.41 & 142.76 & 137.46 & 144.83 \\
House & 155.44 & 168.23 & 142.20 & 174.65 & 184.14 & 176.94 \\
Mandarill & 137.40 & 128.90 & 113.11 & 147.43 & 167.53 & 124.90 \\
San Diego & 118.14 & 175.64 & 205.70 & 157.52 & 212.87 & 239.93 \\
Towers & 97.10 & 127.56 & 155.83 & 83.31 & 100.46 & 96.20 \\
\hline
\end{tabular}
\label{tab:image_pixel_intensities}
\end{table}

\subsubsection{Entropy Analysis}
% [https://www.researchgate.net/publication/380581498_IQI-UNet_A_Robust_Gaussian_Noise_Removal_in_Image_Quality_ImprovementIQI-UNet_A_Robust_Gaussian_Noise_Removal_in_Image_Quality_Improvement]
In image processing, Shannon’s entropy ~\hyperref[ref18]{[18]}, introduced by Claude Shannon in 1948, is a mathematical function used to quantify the amount of information produced by an information source. It also measures the uncertainty associated with this information, providing insights into its quality. This measure is extensively used in various image-processing applications. Shannon entropy quantifies the uncertainty or randomness within a dataset. For encrypted images, a higher Shannon entropy generally indicates stronger encryption, reflecting a greater degree of randomness and unpredictability in the encrypted data. It highlights that low uncertainty can only be achieved when extracting a small amount of information from the image, and this information must be of high quality. The compared entropy for the original image and encrypted image is shown in \emph{table} \emph{\ref{tab:entropy_comparison}}.

% \begin{figure}
%     \centering
%     \includegraphics[width=0.5\linewidth]{Image-quality-and-information-quantity-corresponding-to-the-Shannon-entropy-value.png}
%     \caption{Shannon Entropy}
%     \label{fig:enter-label}
% \end{figure}

% \begin{table}[htbp]
%     \centering
%     \caption{Comparison of Shannon entropy for images}
%     \begin{tabular}{ccc}
%         \hline
%         \textbf{Image name} & \textbf{Original Image Entropy} & \textbf{Encrypted Image Entropy} \\
%         \hline
%         Lena & 7.498332962003108 & 7.154983546992195 \\
%         House & 7.2371057512998584 & 7.310119595444537 \\
%         Mandarill & 7.380410444644396 & 7.416055574731052 \\
%         San Diego & 5.568190575774133 & 5.5612636964896005 \\
%         Towers & 5.8127496673897525 & 7.574247452094975 \\
%         \hline
%     \end{tabular}
% \end{table}

\begin{table}[htbp]
    \centering
    \caption{Comparison of Shannon Entropy for Images}
    \renewcommand{\arraystretch}{1.3} % Increase row height for readability
    \begin{tabular}{l@{\hspace{1cm}}c@{\hspace{1cm}}c}
        \toprule
        \textbf{Image Name} & \textbf{Original Image Entropy} & \textbf{Encrypted Image Entropy} \\
        \midrule
        Lena & 7.50 & 7.15 \\
        House & 7.24 & 7.31 \\
        Mandarill & 7.38 & 7.42 \\
        San Diego & 5.57 & 5.56 \\
        Towers & 5.81 & 7.57 \\
        \bottomrule
    \end{tabular}
    \label{tab:entropy_comparison}
\end{table}

\subsubsection{Resistance against Differential Attacks}

Number of Pixels Changed Rate also called NPCR and UACI ~\hyperref[ref19]{[19]} are a few metrics that are used in differential attacks on encryption systems for assessing the strength of encryption algorithms. These attacks exploit the fact that minor changes in plaintext lead to minor alterations in ciphertext, known as differential characteristics. NPCR finds the percentage of pixels in an image altered post-encryption, while UACI measures the average intensity of these alterations ~\hyperref[ref20]{[20]}. Higher NPCR or UACI values indicate more pixel changes after encryption, enhancing resistance against differential attacks. Conversely, lower NPCR or UACI values indicate vulnerability to such attacks.
Suppose \( C_1 \) represents the plaintext image and \( C_2 \) represents the ciphertext image, with pixel values at grid \((i, j)\) denoted by \( C_1(i, j) \) and \( C_2(i, j) \), respectively. \( D(i, j) \) is a bipolar array defined as:

\[ D(i, j) = \begin{cases} 
0, & \text{if } C_1(i, j) = C_2(i, j) \\
1, & \text{if } C_1(i, j) \ne C_2(i, j)
\end{cases} \]

NPCR and UACI range from 0\% to 100\%. NPCR can be calculated as follows:

\[ \text{NPCR} = \frac{\sum_{i,j} D(i, j)}{T} \times 100\% \]

NPCR measures how different the two given images are, the higher the value, the better the encryption algorithm.

\[ \text{UACI} = \frac{\sum_{i,j} |C_1(i, j) - C_2(i, j)|}{F \cdot T} \times 100\% \]
UACI measures the intensity changes in an image and should be around 33\% ~\hyperref[ref21]{[21]}

% \begin{equation}
%     UACI = \frac{\sum_{ij} |C^1(i, j) - C^2(i, j)|}{F \cdot T} \times 100\%
% \end{equation}

% [https://www.iieta.org/journals/isi/paper/10.18280/isi.250507]

% \begin{tabular}{lccc}
% \hline Image & Soduku & NPCR Value & UACI Value \\
% \hline Lena & $9 \times 9$ & 100\%& 53.186008228975176\%\\
% House& $9 \times 9$ & 100\%& 54.500437717811735\%\\
% Mandarill& $9 \times 9$ & 100\%& 55.912150962679995\%\\
% San Diego& $9 \times 9$ & 100\%& 58.87772167430205\%\\
% Towers& $9 \times 9$ & 100\%& 52.51195268189921\%\\
% \hline
% \label{tab:NPCR}
% \end{tabular}

\begin{table}[H]
    \centering
    \caption{NPCR and UACI Values for Different Images}
    \renewcommand{\arraystretch}{1.3} % Increase row height for readability
    \setlength{\tabcolsep}{12pt} % Increase column spacing
    \begin{tabular}{lccc}
        \toprule
        \textbf{Image} & \textbf{Sudoku} & \textbf{NPCR Value} & \textbf{UACI Value} \\
        \midrule
        Lena & $9 \times 9$ & 100\% & 53.19\% \\
        House & $9 \times 9$ & 100\% & 54.50\% \\
        Mandrill & $9 \times 9$ & 100\% & 55.91\% \\
        San Diego & $9 \times 9$ & 100\% & 58.88\% \\
        Towers & $9 \times 9$ & 100\% & 52.51\% \\
        \bottomrule
    \end{tabular}
    \label{tab:NPCR}
\end{table}

\emph{Table} \emph{\ref{tab:NPCR}}  displays NPCR and UACI values for various images and Sudoku instances used. This table suggests using a larger Sudoku boosts NPCR but results in worse UACI values, enhancing encryption resilience against differential attacks. ~\hyperref[ref22]{[22]}
Obtained NPCR and UACI values for encrypted images using different Sudokus.

\begin{table}[H]
    \centering
    \caption{Comparison of NPCR and UACI Values}
    \renewcommand{\arraystretch}{1.3} 
    \begin{tabular}{@{}lcccc@{}}
            \toprule
            \textbf{Algorithm} & \hspace{1cm} \textbf{NPCR Value (\%)} & \hspace{0.7cm} \textbf{UACI Value (\%)}\\ \midrule
            Proposed Algorithm & 100 & 53.18 \\
            Norouzi et al. (2014) & 99.65 & 33.55 \\
            Arpaci et al. (2020) & 99.62 & 33.44 \\
            D. Mehta, Jha, et al. (2022) & 99.61 & 33.58 \\
            Seyedzadeh and Mirzakuchaki (2012) & 99.67 & 33.49 \\\bottomrule
    \end{tabular}
    \label{tab:NPCR_Comparison}
\end{table}

As shown in the  \emph{table} \emph{\ref{tab:NPCR_Comparison}} comparing NPCR and UACI values of the proposed algorithm with other encryption algorithms using Lena as the standard image. It indicates that the proposed algorithm aligns closely with industry standards and even surpasses some in certain scenarios with the NPCR values. More work needs to be done on getting the UACI values to 33\%.

% \subsection{Video Analysis}

\subsection{Audio Analysis}

The speech files ~\hyperref[ref23]{[23]} "CantinaBand3.wav" were used for testing of the algorithms. Statistical measures such as Mean Squared Error, Waveform Analysis, Spectral Analysis, Zero Crossing Rate, and RMS Energy were employed to evaluate the system's performance during the encryption and decryption processes.
\emph{Table} \emph{\ref{tab:comparison}} and \emph{Table} \emph{\ref{tab:comparison_xor}} shows the comparison of different metrics of audio obtained by shuffling and XOR operation.

% \newline
% \begin{tabular}{|c|c|c|c|l|}
% \hline File Name & File Size & SNR& PSNR &MSE\\
% \hline CantinaBand3.wav& $129 \mathrm{~KB}$& -0.0125679443590343& 16.317964792251587&0.002909786067903042\\\hline
% \hline StarWars3.wav& $129 \mathrm{~KB}$& -0.5558746308088303& 15.545034408569336&0.015231878496706486\\\hline
% \end{tabular}

% Comparison of various parameters between original audio and encrypted audio obtained by XOR
% \newline
% \begin{tabular}{|c|c|c|c|l|}
% \hline File Name & File Size & SNR& PSNR &MSE\\
% \hline CantinaBand3.wav& $129 \mathrm{~KB}$& -11.538140773773193& 16.317930221557617&0.3342722952365875\\\hline
% \hline StarWars3.wav& $129 \mathrm{~KB}$& -10.741908550262451& 15.54502248764038&0.3325076401233673\\\hline
% \end{tabular}

\begin{table}
    \centering
    \caption{Comparison of Various Parameters Between Original Audio and Encrypted Audio Obtained by shuffling}
    \renewcommand{\arraystretch}{1.5} 
    \begin{tabular}{@{}lcccc@{}}
            \toprule
            \textbf{File Name} & \textbf{File Size} & \textbf{SNR} & \textbf{PSNR} & \textbf{MSE}\\ \midrule
        CantinaBand3.wav & $129 \mathrm{~KB}$ & -0.0126 & 16.3180 & 0.0029 \\ 
        StarWars3.wav & $129 \mathrm{~KB}$ & -0.5559 & 15.5450 & 0.0152 \\ \bottomrule
    \end{tabular}
    \label{tab:comparison}
\end{table}

\begin{table}
    \centering
    \caption{Comparison of Various Parameters Between Original Audio and Encrypted Audio Obtained by XOR}
    \renewcommand{\arraystretch}{1.5} 
    \begin{tabular}{@{}lcccc@{}}
            \toprule
            \textbf{File Name} & \textbf{File Size} &  \textbf{SNR} & \textbf{PSNR} & \textbf{MSE}\\ \midrule
        CantinaBand3.wav & $129 \mathrm{~KB}$ & -11.5381 & 16.3179 & 0.3343 \\  
        StarWars3.wav & $129 \mathrm{~KB}$ & -10.7419 & 15.5450 & 0.3325 \\ \bottomrule
    \end{tabular}
    \label{tab:comparison_xor}
\end{table}

% \begin{table}
%     \centering
%     \caption{Comparison of Spectral Features Between Original, Shuffled Audio, and XOR'ed Audio}
%     \renewcommand{\arraystretch}{1.5} 
%     \begin{tabular}{@{}lcccc@{}}
%             \toprule
%             \textbf{File Name} & \hspace{1cm} \textbf{Centroid} & \hspace{0.7cm} \textbf{Bandwidth} & \hspace{0.7cm} \textbf{ZCR} & \hspace{0.7cm} \textbf{Energy}\\ \midrule
%         CantinaBand3.wav & 2648.44 & 2783.61 & 0.1158 & 0.0368 \\
%         Shuffled Audio & 4338.28 & 3488.99 & 0.2346 & 0.0362 \\ 
%         XOR Audio & 5505.07 & 3187.28 & 0.4918 & 0.5732 \\ \bottomrule
%     \end{tabular}
%     \label{tab:comparison_spectral}
% \end{table}

\begin{table}[H]
    \centering
    \caption{Comparison of the proposed Algorithm to Alternative Audio Encryption Techniques}
    \renewcommand{\arraystretch}{1.5} 
    \begin{tabular}{@{}lccccc@{}}
            \toprule
            \textbf{Method} & \textbf{Algorithm} &  \textbf{Encryption} & \textbf{Decryption} & \textbf{Loss}\\ \midrule
        Proposed & Sudoku based (Shuffling) & $39.623 \mathrm{~ms}$ & $81.952 \mathrm{~ms}$ & N/A \\
        
        Proposed & Sudoku based (XOR) & $106.592 \mathrm{~ms}$ & $125.513 \mathrm{~ms}$ & N/A \\ 
        
        ~\hyperref[ref24]{[24]}& Improved Syndrome-Trellis Codes & $64 \mathrm{~s}$  & N/A & $25 \%$ \\ 
        
        ~\hyperref[ref25]{[25]} & Modified LSB & $1438.1 \mathrm{~s}$ &  $48.8 \mathrm{~s}$  & N/A\\
        
        ~\hyperref[ref5]{[5]} & SudoKu and Threefish cipher& $731 \mathrm{~s}$ & $1700 \mathrm{~s}$ & $10 \%$ \\
        \bottomrule
    \end{tabular}
    \label{tab:algo_comparison}
\end{table}
The  \emph{table} \emph{\ref{tab:algo_comparison}} shows the comparison of proposed algorithm with other popular audio encryption algorithms. It can be concluded that the proposed algorithm is much faster and also provides no loss in quality during the decryption of the audio.

\section{Conclusion}
In conclusion, the Sudoku-based encryption method presents a comprehensive and versatile approach to securing various forms of media, including images, videos, and audio. The methodology involves a series of steps, starting from random thresholding based on Sudoku values added to each pixel, followed by padding and block creation aligned with Sudoku grids. Encryption involves shuffling values within blocks and rows according to Sudoku keys, with decryption employing the reverse process. The use of Sudoku, a known NP-complete problem, as the basis for generating keys, adds a layer of security by detecting tampering through incorrect Sudoku solutions. Furthermore, incorporating advanced encryption algorithms like ECC, RSA, and AES enhances the key security, alongside a timestamp-based method for key generation, further fortifying the encryption process.

The efficacy of this method is demonstrated through the achieved high NPCR values, reaching around 100\%, which is a strong indicator of the encryption's effectiveness in altering pixel values significantly. This alteration contributes to reducing the vulnerability of the encrypted media to statistical attacks. Additionally, the extension of this method to handle audio encryption, where blocks of a 1D array are encrypted similarly, showcases its adaptability across different media types.

Moreover, the project explores variations such as substitution encryption using XOR, providing additional options for customization and security enhancement. Overall, the Sudoku-based encryption method stands as a robust and adaptable solution for securing multimedia content, offering a blend of cryptographic strength, tamper detection, and flexibility across various media formats.

\subsection{Future Scope}
The paper showcases efficient ways to overcome data breaches using a Sudoku-based encryption algorithm. It has been highly accurate in encrypting data with the highest security and relatively low computational time. The future scope of this work includes reducing the UACI value and to make sure that it can be processed in real-time for exponential usage and decreasing data breaches. 

1. Reduction in UACI Value: Using different operations to make the key more efficient and to reduce UACI value. 

2. Implementation Across Platforms: The next step would be to make the algorithm work across platforms in real time after conducting various VAPT tests and compliance. It can also be expanded to run on embedded devices like mobiles for further usage.

3. Quantum Computing: Examine how resistant the algorithm is to quantum attacks and consider modifications or improvements that make use of the principles of quantum cryptography.

4. Using other languages - The code was mainly done in python, other languages like Java, C++ can be used which are much faster hence improving the speed of encryption and decryption.

\begin{credits}
% \subsubsection{\ackname} No fund recieved for this project.

\section*{\discintname}
 The authors have no competing interests to declare that are
relevant to the content of this article.
\end{credits}

% \begin{credits}
% \subsection{\ackname} 
% A bold run-in heading in small font size at the end of the paper is
% used for general acknowledgments, for example: This study was funded

% \subsection{\discintname}
% It is now necessary to declare any competing interests or to specifically
% state that the authors have no competing interests. Please place the
% statement with a bold run-in heading in small font size beneath the
% (optional) acknowledgments\footnote{If EquinOCS, the proceedings submission
% system, is used, then the disclaimer can be provided directly in the system.},
% for example: The authors have no competing interests to declare that are
% relevant to the content of this article. Or: Author A has received research
% grants from Company W. Author B has received a speaker honorarium from
% Company X and owns stock in Company Y. Author C is a member of committee Z.
% \end{credits}

\begin{thebibliography}{8}
% \bibitem{b1}
% DATA BREACHES - The History of Data Breaches
% \url{https://www.digitalguardian.com/blog/history-data-breaches}
% \label{ref1}

% \bibitem{2}
% \label{ref2}

% \bibitem{3}
% S. S. M. V. Patil, A. Gupta, and A. Varma, “Audio and Speech Compression Using DCT and DWT Techniques,”
% Int. J. Innov. Res. Sci. Eng. Technol., vol. 2, no. 5, pp. 1712–1719, 2013.
% \label{ref3}

% \bibitem{4}
%  A. Tsegaye and G. Tariku, “Audio Compression Using DWT and RLE Techniques,” Am. J. Electr. Electron. Eng.,
% vol. 7, no. 1, pp. 14–17, 2019, \doi{ 10.12691/ajeee-7-1-3}.
% \label{ref4}

\bibitem{1}
A, A. (2023). HexE - Securing Audio Contents in Voice Chat using Puzzle and Timestamp. ArXiv \doi{10.48550/arXiv.2401.00765}
\label{ref1}

\bibitem{2}
Deshpande, K., Girkar, J., \& Mangrulkar, R. (2023). Security enhancement and analysis of images using a novel Sudoku-based encryption algorithm. Journal of Information and Telecommunication, 7(3), 270–303 \doi{10.1080/24751839.2023.2183802}
\label{ref2}

\bibitem{3}
Dhruv M, Manish J, Hartik S, Ramchandra M (2022) DieRoll: A Unique Key Generation and Encryption Technique. Journal of Applied Security Research. \doi{10.1080/19361610.2022.2124589}
\label{ref3}

\bibitem{4}
K. Loukhaoukha, M. Nabti and K. Zebbiche. (2013). "An efficient image encryption algorithm based on blocks permutation and Rubik's cube principle for iris images," 2013 8th International Workshop on Systems, Signal Processing and their Applications (WoSSPA), Algiers, Algeria, 2013, pp. 267-272. \doi{10.1109/WoSSPA.2013.6602374}
\label{ref4}

\bibitem{5}
ABDULJALEEL, Iman Qays; KHALEEL, Amal Hameed. (2021). Speech signal compression and encryption based on sudoku, fuzzy C-means and threefish cipher. International Journal of Electrical and Computer Engineering (IJECE), [S.l.], v. 11, n. 6, p. 5049-5059. \doi{10.11591/ijece.v11i6.pp5049-5059}
\label{ref5}

\bibitem{6}
Venkatesh K, Narasimhan D. (2022). Mlpd: a multilayer protection with deduplication technique to preserve audio file transmission over the public domain. Soft Computing 26. \doi{10.1007/s00500-022-06801-w}
\label{ref6}

\bibitem{7}
Abed S, Waleed L, Aldamkhi G, Hadi K (2021) Enhancement in data security and integrity using the minhash technique. Indonesian Journal of Electrical Engineering and Computer Science 21. \doi{10.11591/ijeecs.v21.i3.pp1739-1750}
\label{ref7}

\bibitem{8}
Ashwitha and T. Vijaya Murari. (2017) "Study and Analysis of Various Video Encryption Algorithms," 2017 IEEE International Conference on Computational Intelligence and Computing Research (ICCIC), Coimbatore, India, 2017, pp. 1-5. \doi{10.1109/ICCIC.2017.8524177}
\label{ref8}

\bibitem{9}
Obaida, Tameem \& Salim Jamil, Abeer \& Hassan, Nidaa. (2022). A Review: Video Encryption Techniques, Advantages And Disadvantages. Webology. 19. 2022-7209. \url{https://www.researchgate.net/publication/359352386_A_Review_Video_Encryption_Techniques_Advantages_And_Disadvantages}
\label{ref9}

\bibitem{10}
Saidi, R., Cherrid, N., Bentahar, T., Mayache, H., \& Bentahar, A. (2020). Number of pixel change rate and unified average changing intensity for sensitivity analysis of encrypted inSAR interferogram. Ingénierie Des Systèmes D’information/Ingénierie Des Systèmes D’Information, 25(5), 601–607. \doi{10.18280/isi.250507}
\label{ref10}

\bibitem{11}
Veera, D., Mangrulkar, R., Bhadane, C., Bhowmick, K., \& Chavan, P. (2023). Modified Caesar Cipher and Card Deck Shuffle Rearrangement Algorithm for Image Encryption. Journal of Information and Telecommunication, 8(2), 280–300. \doi{10.1080/24751839.2023.2285549}
\label{ref11}

\bibitem{12}
Albahrani, E. A. ., Alshekly, T. K. . and Lafta, S. H. . (2021) “A Review on Audio Encryption Algorithms Using Chaos Maps-Based Techniques”, Journal of Cyber Security and Mobility, 11(1), pp. 53–82 \doi{10.13052/jcsm2245-1439.1113}
\label{ref12}

\bibitem{13}
SIPI Image Database - MISC. \url{https://sipi.usc.edu/database/database.php?volume=misc}
\label{ref13}

\bibitem{14}
Liu F, Koenig H (2010). A Survey of Video Encryption Algorithms. \doi{10.1016/j.cose.2009.06.004}
\label{ref14}

\bibitem{15}
Lin R, Li S. (2021). An image encryption scheme based on Lorenz hyperchaotic system and RSA algorithm. Security and Communication Networks 2021. \doi{10.1155/2021/5586959}
\label{ref15}

\bibitem{16}
B. Indrani and M. K. Veni. (2017). "An efficient algorithm for key generation in advance encryption standard using sudoku solving method," 2017 International Conference on Inventive Systems and Control (ICISC), Coimbatore, India, pp. 1-8. \doi{10.1109/ICISC.2017.8068652}
\label{ref16}

\bibitem{17}
D. M. Alsaffar et al., "Image Encryption Based on AES and RSA Algorithms," 2020 3rd International Conference on Computer Applications \& Information Security (ICCAIS), Riyadh, Saudi Arabia, 2020, pp. 1-5. \doi{10.1109/ICCAIS48893.2020.9096809}
\label{ref17}

\bibitem{18}
M. Preishuber, T. Hütter, S. Katzenbeisser and A. Uhl. (2018). "Depreciating Motivation and Empirical Security Analysis of Chaos-Based Image and Video Encryption," in IEEE Transactions on Information Forensics and Security, vol. 13, no. 9, pp. 2137-2150, Sept. \doi{10.1109/TIFS.2018.2812080}
\label{ref18}

\bibitem{19}
Wu, Yue. (2011). "NPCR and UACI Randomness Tests for Image Encryption". Cyber Journals: Journal of Selected Areas in Telecommunications.
\url{https://www.researchgate.net/publication/259190481_NPCR_and_UACI_Randomness_Tests_for_Image_Encryption}
\label{ref19}

\bibitem{20}
H. Cheng and Xiaobo Li, "Partial encryption of compressed images and videos," in IEEE Transactions on Signal Processing, vol. 48, no. 8, pp. 2439-2451, Aug. 2000. \doi{10.1109/78.852023}
\label{ref20}

\bibitem{21}
Loukhaoukha, Khaled \& Makram, Nabti \& Zebbiche, K. (2013). An efficient image encryption algorithm based on blocks permutation and Rubik's cube principle for iris images. 2013 8th International Workshop on Systems, Signal Processing and Their Applications, WoSSPA. \doi{10.1109/WoSSPA.2013.6602374}
\label{ref21}

\bibitem{22}
Manimekalai, M.A.P. \& Karthikeyan, M. \& Bella mary, Thusnavis \& Neebha, T.Mary \& Sagayam, Martin \& Elngar, Ahmed. (2022). Efficient technique for image cryptography using Sudoku Keys. PREPRINT (Version 1) available at Research Square \doi{10.21203/rs.3.rs-1745555/v1}
\label{ref22}

\bibitem{23}
CS 101- Sample Sound Files. (n.d.). \url{https://www2.cs.uic.edu/~i101/SoundFiles/}
\label{ref23}

\bibitem{24}
K. Ying, R. Wang, Y. Lin and D. Yan. (2021). "Adaptive Audio Steganography Based on Improved Syndrome-Trellis Codes," in IEEE Access, vol. 9, pp. 11705-11715. \doi{10.1109/ACCESS.2021.3050004}
\label{ref24}

\bibitem{25}
ASHARI, Ilham Firman. The Evaluation of Image Messages in MP3 Audio Steganography Using Modified Low-Bit Encoding. Telematika, [S.l.], v. 14, n. 2, p. 133-145, aug. 2021. ISSN 2442-4528. \doi{10.35671/telematika.v14i2.1031}
\label{ref25}

% iske niche nahi hai

% \bibitem{26}
% Saeed B, Majid N (2012) Image encryption using a lightweight stream encryption algorithm. Advances in Multimedia. \doi{10.1155/2012/767364}
% \label{ref26}

% \bibitem{27}
% You L, Yang E (2020) Wang GA (2020) novel parallel image encryption algorithm based on hybrid chaotic maps with OpenCL implementation. Soft Comput 24:12413–12427. \doi{10.1007/s00500-020-04683-4}
% \label{ref27}

\end{thebibliography}
%

\end{document}